\documentclass[
 reprint,
superscriptaddress,
 amsmath,amssymb,
 aps,
floatfix]{revtex4-2}

\usepackage{graphicx}
\usepackage{dcolumn}
\usepackage{bm}
\usepackage{xcolor}
\usepackage{soul}

\begin{document}

\preprint{APS/123-QED}

\title{Comb-referenced Doppler-free spectrometry of the $^{200}$Hg and $^{202}$Hg \\intercombination line at 254 nm}
\author{Stefania Gravina}
 \affiliation{Dipartimento di Matematica e Fisica, Università degli Studi della Campania "Luigi Vanvitelli", 81100, Caserta, Italia}
\author{Naveed A. Chishti}%
\affiliation{%
Dipartimento di Matematica e Fisica, Università degli Studi della Campania "Luigi Vanvitelli", 81100, Caserta, Italia}%
\author{Simona Di Bernardo}
\affiliation{%
Dipartimento di Matematica e Fisica, Università degli Studi della Campania "Luigi Vanvitelli", 81100, Caserta, Italia}%
\author{Eugenio Fasci}
\affiliation{Dipartimento di Matematica e Fisica, Università degli Studi della Campania "Luigi Vanvitelli", 81100, Caserta, Italia}%
\author{Antonio Castrillo}
\affiliation{Dipartimento di Matematica e Fisica, Università degli Studi della Campania "Luigi Vanvitelli", 81100, Caserta, Italia}%
\author{Athanasios Laliotis}
\affiliation{Laboratoire de Physique des Lasers, Université Sorbonne Paris Nord, 93430, Villetaneuse, France}
\author{Livio Gianfrani}%
 \email{Corresponding author: livio.gianfrani@unicampania.it}
\affiliation{%
Dipartimento di Matematica e Fisica, Università degli Studi della Campania "Luigi Vanvitelli", 81100, Caserta, Italia}%

\date{\today}

\begin{abstract}
We report on precision spectroscopy of the 6s$^2$ $^1$S$_0\to$6s6p $^3$P$_1$ intercombination line of mercury in the deep ultraviolet, by means of a frequency-comb referenced, wavelength-modulated, saturated absorption technique. This method allowed us to perform sub-Doppler investigations with an absolute frequency axis at 254 nm, while ensuring a relatively high signal-to-noise ratio. The absolute line center frequencies of the $^{200}$Hg and $^{202}$Hg bosonic isotopes were measured with a global uncertainty of 8 and 15 kHz (namely, 6.8$\times$10$^{-12}$ and 1.3$\times$10$^{-11}$, in relative terms), respectively, the statistical and systematic components being significantly reduced as compared to past determinations. This remarkable result was achieved also thanks to an in-depth study of the AC stark effect. Furthermore, we found the most accurate $^{200}$Hg-$^{202}$Hg isotope shift ever obtained before, namely, $5295570\pm15_{stat}\pm8_{syst}$  kHz.  
\end{abstract}

\maketitle


In recent years, renewed interest has emerged in the isotopic shift of atomic transitions. In fact, isotope shift spectroscopy in heavy atoms can be used to look for new forces beyond the standard model, as well as to determine basic properties of atomic nuclei \cite{Frugiuele2017}. More particularly, when collecting spectroscopic data on at least four isotopes, it is possible to constrain new physics scenarios by testing very accurately the linearity of King plots \cite{Flambaum2018}. The King plot analysis was extensively used in the 80s and 90s to investigate the field shift, which originates from the different nuclear volumes of the various isotopes of a given element \cite{King2013,Rayman89,Gianfrani1990}. Nevertheless, there might be nonlinear corrections to the King plots that can be justified already in the standard model framework. This is the case of the nuclear polarizability contribution, which can lead to a significant deviation from linearity. Very recently, the clear nonlinearity in the joint Yb/Yb$^+$ King-plot analysis has been mostly accounted for by the deformation of the Yb nuclei \cite{Figueroa2022,Hur2022}. In this regard, it has been argued that a direct comparison between theory and experiment on a single transition for a single pair of isotopes can yield more stringent bounds with the same amount of theoretical effort \cite{Debierre2022}.

A few years ago, absolute center frequencies and isotope shifts have been determined for the Hg 6s$^2$ $^1$S$_0\to$6s6p $^3$P$_1$ line with an accuracy of a few hundred kHz, namely, more than one order of magnitude better than any previous measurement \cite{Witkowski2019}. With its seven stable isotopes, four of which having zero nuclear spin ($^{198}$Hg, $^{200}$Hg, $^{202}$Hg, and $^{204}$Hg), mercury is a quite interesting atom for fundamental tests and measurements \cite{Safronova2018}. It is a good candidate for the development of a primary frequency reference, due to some favorable characteristics, such as its sensitivity to blackbody radiation, which is lower than that of Sr and Yb \cite{McFerran2012}. The intercombination transition plays a key role in the search for a permanent electric dipole moment in the $^1$S$_0$ electronic ground state of the $^{199}$Hg atom, the spectral line being involved in a pump-probe process based upon spin-polarization \cite{Graner2016}. As for many transitions from the ground state ns$^2$ to the $nsnp$ configurations of two-electron atoms or ions, the Hg intercombination line is a highly sensitive probe of possible long-term variations of the fine-structure constant \cite{Angstmann2004}. 
An ongoing experiment in Caserta employs the Hg intercombination line at 253.7 nm for temperature metrology, by using comb-calibrated Doppler width spectrometry \cite{Clivati2020, Gravina2023}. This latter is a promising primary method for the practical realization of the new kelvin definition, which is in force since 20 May 2019 \cite{Machin}. 
In this technique, once the Doppler width is retrieved from the precise observation of the absorption lineshape in a gas at the thermodynamic equilibrium, the absolute temperature can be determined by inverting the well-known expression of the Doppler width, provided that the line center frequency is accurately known \cite{Gianfrani2016linking}.  

In this framework, the present manuscript has a twofold objective: to measure the center frequency of the Hg intercombination line at the precision level of 10 kHz; to improve significantly the current knowledge of the $^{200}$Hg-$^{202}$Hg isotope shift of this spectral line. To pursue these goals, we developed a rather advanced apparatus capable of Doppler-free precision spectroscopic measurements with a wavelength-modulation (WM) technique, by using low-power coherent radiation in the deep-ultraviolet region.
\begin{figure*}
   \centering
\includegraphics[scale=0.70]{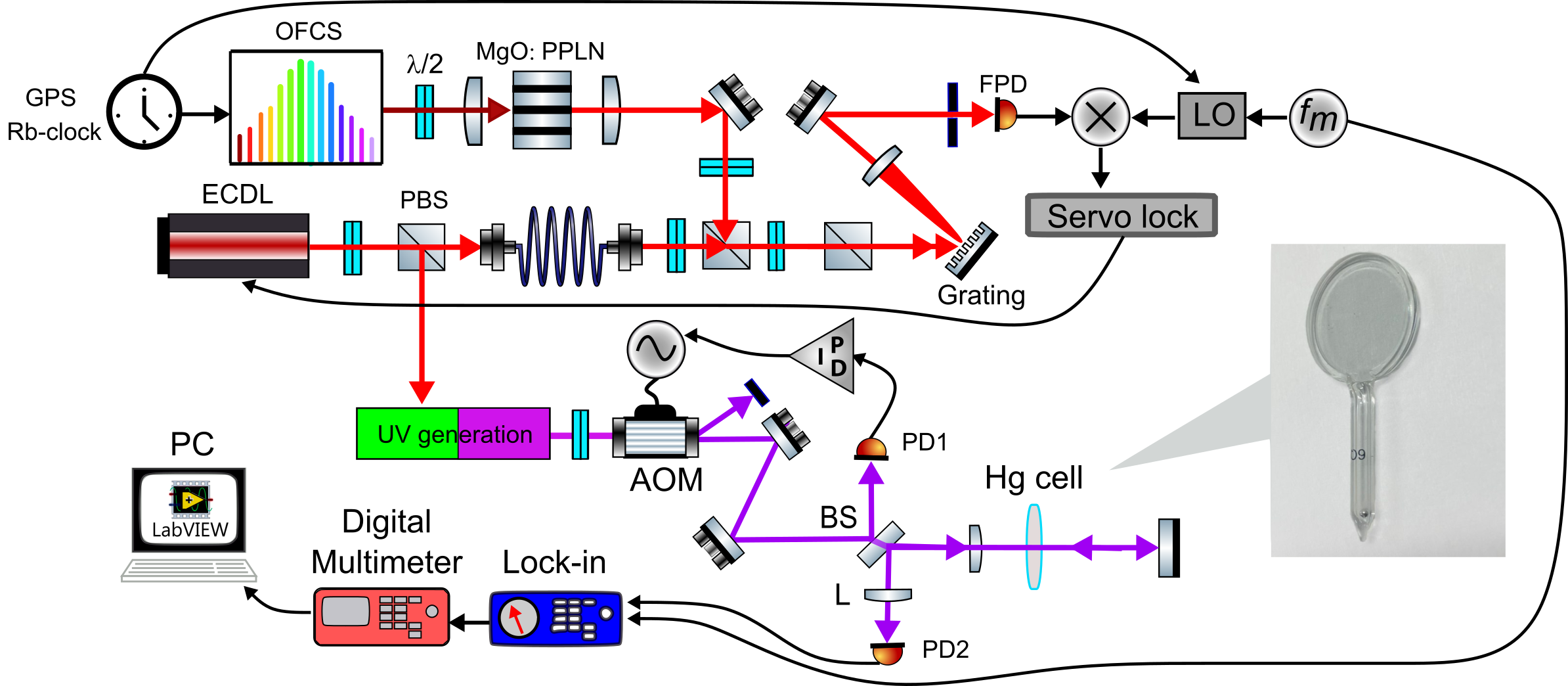}
\caption{Sketch of the experimental setup. ECDL stands for external-cavity diode laser; OFCS, optical frequency comb synthesizer; MgO:PPLN, periodically poled MgO-doped LiNbO$_3$ crystal; PD, photodiode; FPD, fast photodiode; LO, local oscillator; PBS, polarizing beam splitter; BS, 50$\%$ beam splitter; L, plano-convex lens.}
    \label{apparato}
\end{figure*}
The spectrometer is schematically depicted in Fig. \ref{apparato}.
As extensively described in Refs. \cite{Clivati2020, Gravina2022}, the laser source consists of a double-stage duplication of a 1014.8 nm external-cavity diode laser (ECDL). 
After spatial filtering by means of a 200-$\mu$m pinhole, the UV radiation at the wavelength of 253.7 nm passes through an acousto-optic modulator (AOM), which has a twofold function: first, it is used to finely change the UV power of the zero-order that is used for the spectroscopic experiment; second, it is the actuator of an intensity control feedback loop, thereby stabilizing the UV power at the level of one part in 10$^4$ within a few tens of minutes, in the baseband region. To this aim, a portion of the UV beam is monitored by a first silicon carbide (SiC) photodetector. The remaining portion of the UV radiation (whose power can go up to $\sim$10 $\mu$W, depending on the radio-frequency power driving the AOM) is directed towards a 1-mm-long mercury vapour cell. Sealed at the two ends by a pair of Spectrosil far-UV quartz windows, the cell has a cylindrical geometry with a 2-cm diameter and it is equipped with a small reservoir in the middle of its length.
As guaranteed by the manufacturer (Precision Glassblowing Inc.), the cell is baked at 425 $^{\circ}$C for a minimum of 24 hours in an ultra-high vacuum environment prior to any introduction of mercury in the reservoir. This process ensures ultrapure Hg vapors in equilibrium with the liquid phase inside the cell.
The reservoir is immersed into a home-made liquid-bath thermostat based on Peltier cells, whose temperature can be adjusted from 0 to 50 $^{\circ}$C. So doing, it is possible to tune the vapor pressure in the thin cell roughly from 2$\times$10$^{-2}$ to 2 Pa. 

We implemented the conventional pump–probe scheme in which the traveling laser beam, after passing through the thin cell, is retroreflected towards the cell by a plane mirror. To ensure the precise alignment and overlap of the two beams, a pair of pinholes are positioned along the beam path. Care was taken to gently focus the pump beam in the center of the cell. The reflected probe beam returns to the beam splitter and subsequently is focused on a second SiC photodetector.

The near-infrared ECDL is tightly locked to an optical frequency comb synthesizer (MENLO Systems, FC1500-250-WG) based on a mode-locked, femtosecond, erbium-doped fiber laser at 1560 nm and stabilized against a GPS-disciplined Rb clock. Details are provided in the Supplemental Material. The absolute UV frequency can be easily retrieved from the following equation:
\begin{equation}
 f_{UV}=4 \times ( f_{beat} + m f_{rep} + 2f_{ceo}),
    \label{fUV}
\end{equation}
where $f_{beat}$ is the beat note frequency at 1015 nm, $m$ is the comb tooth order, $f_{rep}$ and $f_{ceo}$ are the comb repetition rate (=250 MHz) and the carrier-envelope offset frequency (=20 MHz), respectively.
The comb order was unambiguously determined by measuring the ECDL wavelength  in the near-infrared with a 9-digit wavemeter.

High sensitivity detection of the relatively small Lamb-dip was achieved
by using a WM technique, which is inspired by the developments of Gambetta $et$ $al.$ in the field of mid-infrared frequency-comb referenced molecular spectroscopy \cite{Gambetta}. To this aim, the LO frequency is dithered at $f_m$=220 Hz. Thanks to the action of the locking loop, the sinusoidal modulation is effectively transferred to the ECDL frequency and, consequently, to the UV radiation so as to set the modulation index to about 1.
An analog lock-in amplifier was used to perform the first-harmonic synchronous detection of the probe beam intensity, thus leading to a dispersive signal in coincidence with the Lamb-dip. Calibrated frequency scans across the intercombination line were obtained by finely tuning the $f_{rep}$ frequency while the ECDL was locked to the comb. 
A $6\frac{1}{2}$-digit voltmeter, connected to a personal computer, is used for spectral acquisitions, as governed by a LabVIEW code that also controls $f_{rep}$.

\begin{figure}
   \centering
\includegraphics[width=\linewidth]{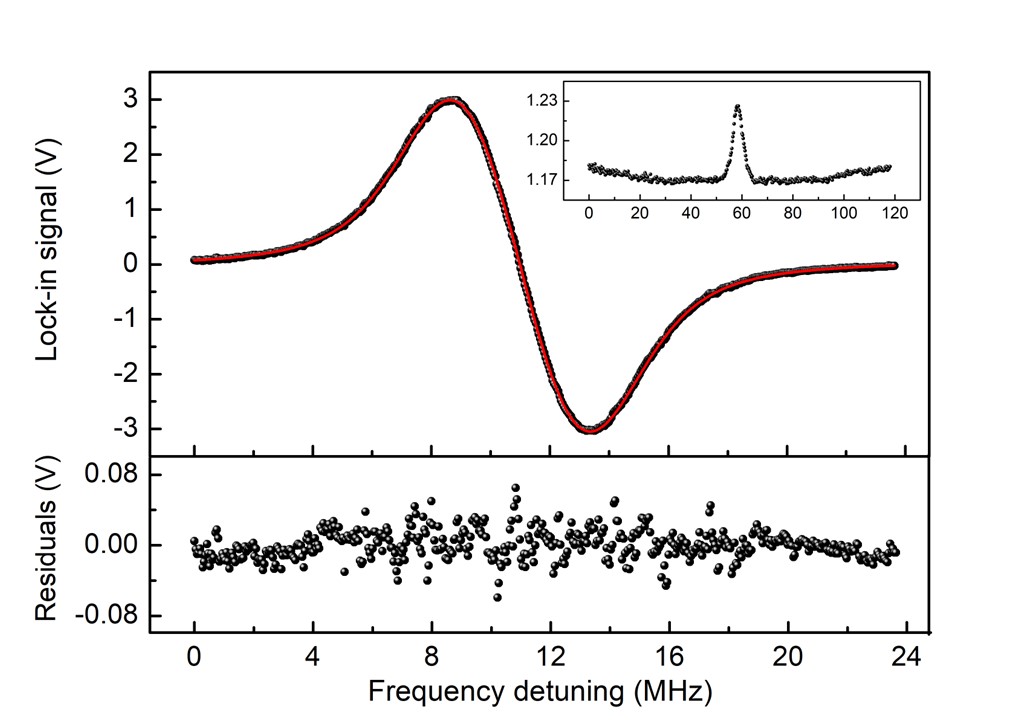}
    \caption{The top panel shows the dispersive $^{200}$Hg sub-Doppler spectrum, as obtained by means of the wavelength-modulated technique at a vapor pressure of $\sim$0.24 Pa. For comparison, an example of Lamb-dip spectrum is reported in the inset. The red line represents the best-fit curve, while the bottom plot gives the residuals. The SNR amounts to $\sim$370. The saturation intensity was calculated to be $\sim$110 mW/cm$^2$, while the UV radiation intensity in the beam waist was $\sim$15 mW/cm$^2$.
\label{spettro_derivata}}
\end{figure}

An example of sub-Doppler spectrum for the $^{200}$Hg 6s$^2$ $^1$S$_0\to$6s6p $^3$P$_1$ transition is shown in Fig. \ref{spettro_derivata}. Compared to direct detection of absorption, which leads to Lamb-dip spectra similar to the one shown in the inset of Fig. \ref{spettro_derivata}, the WM technique ensures background flattening in conjunction to a higher signal-to-noise ratio (SNR), circumstance that allows one to retrieve the line center frequency more accurately. This spectrum consists of 500 points that are acquired in 73 s. We operate in the weak saturation regime, the Lamb-dip contrast ranging between 3 and 5$\%$, while the single-pass linear absorption amounts to $\sim$54$\%$. A refined line fitting is a key factor to accurately retrieve the absolute center frequency. It is well known that wavelength modulation leads to lock-in signals that are proportional to various Fourier coefficients of the modulated lineshape function, depending on the order of the harmonic that is detected. We adopted as a lineshape model the wavelength-modulated Voigt function, as described in Ref. \cite{Westberg2012}. Details about the lineshape model are provided in the Supplemental Material. Fit residuals exhibit a root-mean-square value of about 16 mV, thus demonstrating the excellent agreement between the experimental data and the lineshape model, as shown in the bottom plot of Fig. \ref{spettro_derivata}. The spectral analysis gives a full width at half maximum (FWHM) of the sub-Doppler line of about 5 MHz, which includes the natural width of the line ($\approx$1.3 MHz), Zeeman and collisional broadening, as well as the transit-time broadening ($\approx$320 kHz). It is worth noting that the observed width is close to the expected one ($\approx$6 MHz). The slight discrepancy is probably due to an overestimation of the calculated Zeeman broadening effect.

The experimental strategy is to retrieve one center frequency value from a pair of round-trip scans, to cancel out the frequency shifts induced by the limited detection bandwidth \cite{Rohart2014, Rohart2017}. Additional information on this effect are provided in the Appendix A. Moreover, Lamb-dip measurements were repeated as a function of the incident UV power to take into account the possible occurrence of the AC Stark shift. In our experiment, the UV power hitting the cell was varied from 1 up to 10 $\mu$W, which gives an intensity in the beam waist, $I_0$, approximately between 3 and 30 mW/cm$^2$.
These values were determined by using a UV beam profiler in conjunction to a calibrated SiC photodiode. 
\begin{figure} 
   \centering
\includegraphics[width=\linewidth]{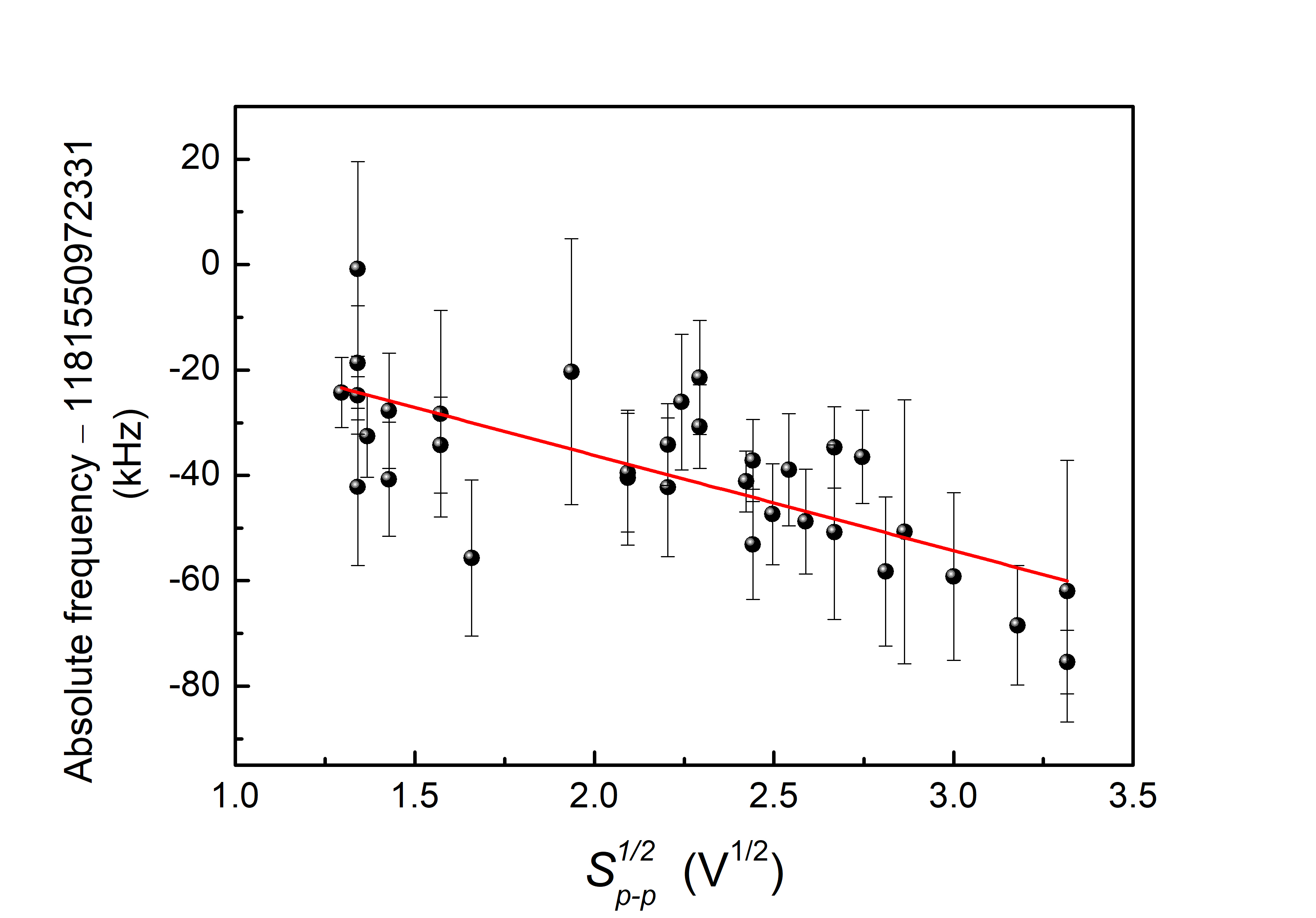}
\caption{Line centers as a function of the square root of the peak-to-peak signal. Each point is the mean value of the center frequencies from two consecutive scans at a certain UV power. Vertical bars result from error propagation of line fitting uncertainties and subsequent multiplication by a factor of 3. This is done in order to be very prudent in quoting the statistical uncertainty of the retrieved parameters. The red solid line is the weighted linear fit of the data. 
\label{Stark}}
\end{figure}
In Fig. \ref{Stark}, we report the results of this investigation for the $^{200}$Hg isotope. The center frequencies are plotted as a function of the square root of the peak-to-peak signal, $S_{p-p}$. This latter can be retrieved from the fit of wavelength-modulated sub-Doppler spectra. Furthermore, $S_{p-p}$ is proportional to the Lamb-dip amplitude, which in turn increases with $I_0^2$, in the weak saturation regime. Fig. \ref{Stark} shows that our measurements are clearly affected by a negative AC Stark shift. A linear regression of the data gives the zero-intensity center frequency, equal to 1181550972331(6) kHz, the reported uncertainty being of pure statistical nature at 1$\sigma$.
The measurement procedure was also applied to the $^{202}$Hg line, resulting in an absolute center frequency of 1181545676761(14) kHz.
After proper calibration of the peak-to-peak signal, the AC Stark effect was quoted to be -2.8$\pm$0.5 kHz/mW/cm$^2$ and -2.2$\pm$0.8 kHz/mW/cm$^2$, respectively for $^{200}$Hg and $^{202}$Hg. The former differs significantly from the estimate of Ref. \cite{Witkowski2019}, while the latter is in good agreement. 
It should be noted that, in the two-level atom approximation, no frequency shift is expected from the resonant UV radiation. Nevertheless, light shift contributions can originate by other electric-dipole transitions involving one of the two states of the intercombination transition, whose frequencies are close to the incident UV radiation. Using time-dependent second-order perturbation theory, energy shifts due to nonresonant light can be estimated \cite{Budker2006}. We found that the shift of the intercombination line is mostly due to the negative energy shift of the 6s6p $^3$P$_1$ state, this latter showing a strong dipole coupling with highly excited states, including 6s9s $^3$S$_1$ and 6s8d $^3$D$_{0,1,2}$.
\begin{table}[b]
\caption{\label{budget-errori}
Uncertainty budget related to the line center frequency measurements for the $^{200}$Hg and $^{202}$Hg isotopes.
}
\begin{ruledtabular}
\begin{tabular}{lcr}
\textrm{Contribution}&
\textrm{Type A}&
\textrm{Type B}\\
&(kHz)&(kHz)\\
\colrule
Statistical uncertainty& 6$-$14\\
  Frequency calibration && 3 \\
       Pressure shift & & 5  \\
       2$^{nd}$-order Doppler shift & & $<$0.2  \\
   Overall combined uncertainty & \multicolumn{2}{c}{8$-$15}\\
\end{tabular}
\end{ruledtabular}
\end{table}

Table \ref{budget-errori} presents the complete uncertainty budget. 
The OFCS contributes with an uncertainty of $\sim$3 kHz, arising from the stability of the GPS-disciplined Rb clock, which amounts to 2.5$\times$10$^{-12}$ at 1 s. 
We quantified the contribution of the pressure shift, also taking into account residual gases that may form in the sealed cell, namely, helium, water and molecular hydrogen. Diffusion from ambient air may lead to a He partial pressure into the quartz cell of up to $\sim$0.5 Pa. From the pressure shifting coefficient due to He-Hg collisions, as reported in Ref. \cite{Lewis1980}, we retrieved an upper limit to the shift of +5 kHz. Instead, the estimated coefficient of the Hg-Hg collision-induced frequency shift results to be -22 kHz/Pa, in sub-Doppler experiments \cite{Witkowski2019}. This latter produces a negative shift of -5.3 kHz. As for the other gases, H$_2$O and H$_2$, we expect partial pressures much smaller than 1 Pa, due to the cell manufacturing process. This is confirmed by the relatively narrow Lamb-dips that are observed. Since the joint effect of the various collision partners partially cancels out, a conservative estimate of the uncertainty contribution coming from the overall pressure shift can be 5 kHz at 1$\sigma$. The 2$^{nd}$-order Doppler shift was quoted to be about -160 Hz. The recoil effect in the case of two counter-propagating beams leads to a doublet whose frequency separation is $\sim$30 kHz, namely, much smaller that the width of the sub-Doppler feature. Due to its symmetry, the unresolved doublet is not expected to produce a shift of the center frequency \cite{Petersen}. Similarly, possible shifts resulting from Zeeman, gas-lens and wavefront-curvature-induced residual Doppler effects are negligible.   
The overall uncertainties are quoted to be 8 and 15 kHz, respectively for $^{200}$Hg and $^{202}$Hg. Consequently, the level of knowledge is improved by more than a factor of 20 for $^{200}$Hg and a factor of 7 for $^{202}$Hg, compared to the most recent determinations \cite{Witkowski2019}.   
From the absolute frequency measurements given above, it is possible to determine the shift between the $^{200}$Hg and $^{202}$Hg isotopes, which amounts to $5295570\pm15_{stat}\pm8_{syst}$  kHz.
Our determination should be compared with the one resulting from the data of Ref. \cite{Witkowski2019}, namely, $5295413\pm110_{stat}\pm180_{syst}$  kHz. The two values agree satisfactorily, their difference being 157 kHz. 

In conclusion, we have improved significantly the current knowledge of the absolute frequency of the mercury intercombination line in the deep-UV, using comb-calibrated wavelength-modulated saturated absorption spectroscopy, a powerful technique that allowed us to achieve a relative precision of 6.8$\times$10$^{-12}$ and 1.3$\times$10$^{-11}$, respectively for the $^{200}$Hg and $^{202}$Hg isotopes. As a result, the $^{200}$Hg-$^{202}$Hg isotope shift could be measured more precisely and accurately, the statistical and systematical uncertainties being reduced by a factor of $\sim$7 and $\sim$23, respectively, compared to the recent experiment of Ref. \cite{Witkowski2019}. It would be interesting to make a comparison with accurate theoretical calculations. In this regard, our study can stimulate the extension to Hg isotopes of relativistic multiconfiguration Dirac-Hartree-Fock calculations, recently performed for the ytterbium intercombination line \cite{McFerran}, thus paving the way to experimental and theoretical investigations of quantum electrodynamics and nuclear size effects in Hg atoms for testing fundamental physical theories. Furthermore, the application of our method to ultra-thin Hg cells of sub-$\mu$m thickness would provide the possibility to look for Casimir-Polder interactions with a non-hydrogen-like atom or study atom-wall collisions in mercury vapors.

\begin{acknowledgments}
This research was partially funded by Università degli Studi della Campania Luigi Vanvitelli under the program ''Progetto Giovani'', Call 2022.
\end{acknowledgments}

\begin{appendix}
\appendix
\section{The finite bandwidth effect}
\begin{figure} 
   \centering
\includegraphics[width=\linewidth]{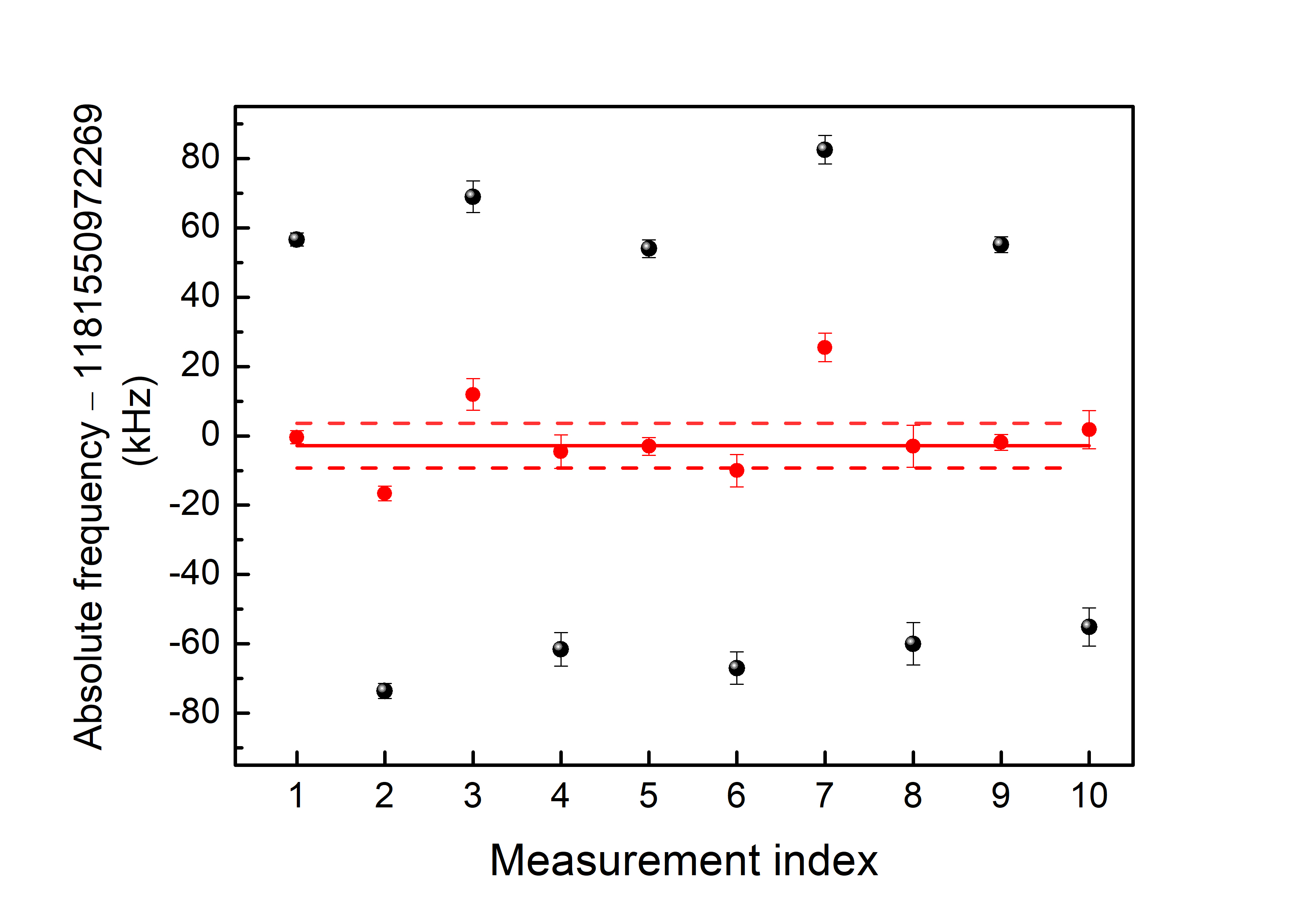}
    \caption{Finite bandwidth effect. Absolute line center frequencies for the $^{200}$Hg isotope at 24$^{\circ}$C are shown as black points. The red points are the corrected frequencies. 
\label{confronto}}
\end{figure}

Fig. \ref{confronto} shows the results obtained from the spectral analysis of 10 repeated acquisitions of the sub-Doppler feature at the liquid mercury temperature of 24$^{\circ}$C, with an incident power of $\sim$9 $\mu$W. 
The line center frequencies that are retrieved from scans with a positive sweep speed, $\dot{\nu}$ = d$\nu$/dt, are clearly shifted with respect to those resulting from scans with $\dot{\nu}<$0. The former data are labeled in Fig. \ref{confronto} with odd acquisition numbers, while the latter present an even acquisition index. 
This is a systematic deviation that is due to the finite bandwidth of the detection electronics with which spectra are recorded, an effect that is often underestimated \cite{Rohart2014}. It is worth noting that the effect is clearly measurable despite the fact that no line asymmetry is observed. 
The sub-Doppler spectra are acquired with a sweep speed of $\pm$0.326 MHz/s and a lock-in amplifier time constant, $\tau_{D}$, of 100 ms.
Each frequency step, $\Delta \nu$, has a time duration, $\Delta t$, of 145 ms. The lock-in amplifier, operating in the -12 dB/octave roll-off mode, can be conceptualized as two consecutive first-order filters with identical characteristics \cite{Rohart2014}. 
In this mode of operation, the filter quality factor, is equal to 1/2. According to the description of Ref. \cite{Rohart2017}, which applies to a step-by-step frequency-sweeping mode, it is possible to define a frequency constant given by $\nu_{D}=\theta\Delta \nu$, $\theta$ being the ratio $\tau_{D}/\Delta t$ that is equal to $\sim$0.66 in our operation conditions. It is found that the retrieved center frequency, denoted as $\nu_{0fit}$, deviates from the true value according to the equation \cite{Rohart2017}:
\begin{equation}
   \nu_{0fit}-\nu_0= \dfrac{exp(-\dfrac{1-2\eta}{2\theta})}{sinh(\dfrac{1}{2\theta})} \times A\times \Delta\nu/2, 
\end{equation}
where $\eta$ is the dead-time constant that is defined as $\dfrac{\Delta t-\Delta t_a}{\Delta t}$, namely, the ratio between the technical dead-time occurring between signal acquisition (over a time span $\Delta t_a$) and next frequency change and the step duration.
The quantity $A$ is a function of $\eta$ and $\theta$, defined as follows:
\begin{equation}
A(\eta, \theta) = 1+\dfrac{1-2\eta}{2\theta}+\dfrac{1}{2\theta}coth(\dfrac{1}{2\theta}). 
\end{equation} 
Since the dead time for each frequency step is $\sim$50 ms, the frequency shift amounts to $\pm$57 kHz (depending on the sign of $\Delta\nu$), which is in good agreement with the observed systematic deviations. Therefore, as a side result of our work, we can confirm the validity of the model of Ref. \cite{Rohart2017}, as demonstrated by the red points of Fig. \ref{confronto}, for which the frequency shift has been removed. Here, the zero of the vertical scale is set in coincidence with the mean value of the uncorrected frequencies (black points), while the solid red line gives the weighted mean of the corrected frequencies (red points), the dash-dot lines representing the 2-$\sigma$ confidence interval. We used the corrected uncertainty on the weighted mean as a result of a mutual consistency test, thus following the recommendations of the \textit{Bureau International des Poids et Mesures} (BIPM) \cite{BIPM}. The mean value falls well within this range. On the other hand, Eq. (A1) tells us that the systematic deviations cancel out if we take the mean value of the center frequencies resulting from two consecutive scans, the former characterized by $\Delta \nu>0$, the latter with $\Delta \nu<0$. This is the reason why it is essential to obtain the center frequency value from two consecutive scans.
\end{appendix}

\nocite{*}


\begin{thebibliography}{27}%
\makeatletter
\providecommand \@ifxundefined [1]{%
 \@ifx{#1\undefined}
}%
\providecommand \@ifnum [1]{%
 \ifnum #1\expandafter \@firstoftwo
 \else \expandafter \@secondoftwo
 \fi
}%
\providecommand \@ifx [1]{%
 \ifx #1\expandafter \@firstoftwo
 \else \expandafter \@secondoftwo
 \fi
}%
\providecommand \natexlab [1]{#1}%
\providecommand \enquote  [1]{``#1''}%
\providecommand \bibnamefont  [1]{#1}%
\providecommand \bibfnamefont [1]{#1}%
\providecommand \citenamefont [1]{#1}%
\providecommand \href@noop [0]{\@secondoftwo}%
\providecommand \href [0]{\begingroup \@sanitize@url \@href}%
\providecommand \@href[1]{\@@startlink{#1}\@@href}%
\providecommand \@@href[1]{\endgroup#1\@@endlink}%
\providecommand \@sanitize@url [0]{\catcode `\\12\catcode `\$12\catcode `\&12\catcode `\#12\catcode `\^12\catcode `\_12\catcode `\%12\relax}%
\providecommand \@@startlink[1]{}%
\providecommand \@@endlink[0]{}%
\providecommand \url  [0]{\begingroup\@sanitize@url \@url }%
\providecommand \@url [1]{\endgroup\@href {#1}{\urlprefix }}%
\providecommand \urlprefix  [0]{URL }%
\providecommand \Eprint [0]{\href }%
\providecommand \doibase [0]{https://doi.org/}%
\providecommand \selectlanguage [0]{\@gobble}%
\providecommand \bibinfo  [0]{\@secondoftwo}%
\providecommand \bibfield  [0]{\@secondoftwo}%
\providecommand \translation [1]{[#1]}%
\providecommand \BibitemOpen [0]{}%
\providecommand \bibitemStop [0]{}%
\providecommand \bibitemNoStop [0]{.\EOS\space}%
\providecommand \EOS [0]{\spacefactor3000\relax}%
\providecommand \BibitemShut  [1]{\csname bibitem#1\endcsname}%
\let\auto@bib@innerbib\@empty
\bibitem [{\citenamefont {Frugiuele}\ \emph {et~al.}(2017)\citenamefont {Frugiuele}, \citenamefont {Fuchs}, \citenamefont {Perez},\ and\ \citenamefont {Schlaffer}}]{Frugiuele2017}%
  \BibitemOpen
  \bibfield  {author} {\bibinfo {author} {\bibfnamefont {C.}~\bibnamefont {Frugiuele}}, \bibinfo {author} {\bibfnamefont {E.}~\bibnamefont {Fuchs}}, \bibinfo {author} {\bibfnamefont {G.}~\bibnamefont {Perez}},\ and\ \bibinfo {author} {\bibfnamefont {M.}~\bibnamefont {Schlaffer}},\ }\bibfield  {title} {\bibinfo {title} {Constraining new physics models with isotope shift spectroscopy},\ }\href {https://doi.org/10.1103/PhysRevD.96.015011} {\bibfield  {journal} {\bibinfo  {journal} {Phys. Rev. D}\ }\textbf {\bibinfo {volume} {96}},\ \bibinfo {pages} {015011} (\bibinfo {year} {2017})}\BibitemShut {NoStop}%
\bibitem [{\citenamefont {Flambaum}\ \emph {et~al.}(2018)\citenamefont {Flambaum}, \citenamefont {Geddes},\ and\ \citenamefont {Viatkina}}]{Flambaum2018}%
  \BibitemOpen
  \bibfield  {author} {\bibinfo {author} {\bibfnamefont {V.~V.}\ \bibnamefont {Flambaum}}, \bibinfo {author} {\bibfnamefont {A.~J.}\ \bibnamefont {Geddes}},\ and\ \bibinfo {author} {\bibfnamefont {A.~V.}\ \bibnamefont {Viatkina}},\ }\bibfield  {title} {\bibinfo {title} {{Isotope shift, nonlinearity of King plots, and the search for new particles}},\ }\href {https://doi.org/10.1103/PhysRevA.97.032510} {\bibfield  {journal} {\bibinfo  {journal} {Phys. Rev. A}\ }\textbf {\bibinfo {volume} {97}},\ \bibinfo {pages} {032510} (\bibinfo {year} {2018})}\BibitemShut {NoStop}%
\bibitem [{\citenamefont {King}(2013)}]{King2013}%
  \BibitemOpen
  \bibfield  {author} {\bibinfo {author} {\bibfnamefont {W.~H.}\ \bibnamefont {King}},\ }\href@noop {} {\emph {\bibinfo {title} {Isotope shifts in atomic spectra}}}\ (\bibinfo  {publisher} {Springer Science \& Business Media},\ \bibinfo {year} {2013})\BibitemShut {NoStop}%
\bibitem [{\citenamefont {Rayman}\ \emph {et~al.}(1989)\citenamefont {Rayman}, \citenamefont {Aminoff},\ and\ \citenamefont {Hall}}]{Rayman89}%
  \BibitemOpen
  \bibfield  {author} {\bibinfo {author} {\bibfnamefont {M.~D.}\ \bibnamefont {Rayman}}, \bibinfo {author} {\bibfnamefont {C.~G.}\ \bibnamefont {Aminoff}},\ and\ \bibinfo {author} {\bibfnamefont {J.~L.}\ \bibnamefont {Hall}},\ }\bibfield  {title} {\bibinfo {title} {Precise laser frequency scanning using frequency-synthesized optical frequency sidebands: application to isotope shifts and hyperfine structure of mercury},\ }\href@noop {} {\bibfield  {journal} {\bibinfo  {journal} {J. Opt. Soc. Am. B}\ }\textbf {\bibinfo {volume} {6}},\ \bibinfo {pages} {539} (\bibinfo {year} {1989})}\BibitemShut {NoStop}%
\bibitem [{\citenamefont {Gianfrani}\ \emph {et~al.}(1990)\citenamefont {Gianfrani}, \citenamefont {Sasso}, \citenamefont {Tino},\ and\ \citenamefont {Inguscio}}]{Gianfrani1990}%
  \BibitemOpen
  \bibfield  {author} {\bibinfo {author} {\bibfnamefont {L.}~\bibnamefont {Gianfrani}}, \bibinfo {author} {\bibfnamefont {A.}~\bibnamefont {Sasso}}, \bibinfo {author} {\bibfnamefont {G.~M.}\ \bibnamefont {Tino}},\ and\ \bibinfo {author} {\bibfnamefont {M.}~\bibnamefont {Inguscio}},\ }\bibfield  {title} {\bibinfo {title} {Experimental indication of a nuclear volume contribution to the isotope shift of atomic oxygen},\ }\href@noop {} {\bibfield  {journal} {\bibinfo  {journal} {Optics Communications}\ }\textbf {\bibinfo {volume} {78}},\ \bibinfo {pages} {158} (\bibinfo {year} {1990})}\BibitemShut {NoStop}%
\bibitem [{\citenamefont {Figueroa}\ \emph {et~al.}(2022)\citenamefont {Figueroa}, \citenamefont {Berengut}, \citenamefont {Dzuba}, \citenamefont {Flambaum}, \citenamefont {Budker},\ and\ \citenamefont {Antypas}}]{Figueroa2022}%
  \BibitemOpen
  \bibfield  {author} {\bibinfo {author} {\bibfnamefont {N.~L.}\ \bibnamefont {Figueroa}}, \bibinfo {author} {\bibfnamefont {J.~C.}\ \bibnamefont {Berengut}}, \bibinfo {author} {\bibfnamefont {V.~A.}\ \bibnamefont {Dzuba}}, \bibinfo {author} {\bibfnamefont {V.~V.}\ \bibnamefont {Flambaum}}, \bibinfo {author} {\bibfnamefont {D.}~\bibnamefont {Budker}},\ and\ \bibinfo {author} {\bibfnamefont {D.}~\bibnamefont {Antypas}},\ }\bibfield  {title} {\bibinfo {title} {{Precision determination of isotope shifts in Ytterbium and implications for new physics}},\ }\href {https://doi.org/10.1103/PhysRevLett.128.073001} {\bibfield  {journal} {\bibinfo  {journal} {Phys. Rev. Lett.}\ }\textbf {\bibinfo {volume} {128}},\ \bibinfo {pages} {073001} (\bibinfo {year} {2022})}\BibitemShut {NoStop}%
\bibitem [{\citenamefont {Hur}\ \emph {et~al.}(2022)\citenamefont {Hur}, \citenamefont {Aude~Craik}, \citenamefont {Counts}, \citenamefont {Knyazev}, \citenamefont {Caldwell}, \citenamefont {Leung}, \citenamefont {Pandey}, \citenamefont {Berengut}, \citenamefont {Geddes}, \citenamefont {Nazarewicz}, \citenamefont {Reinhard}, \citenamefont {Kawasaki}, \citenamefont {Jeon}, \citenamefont {Jhe},\ and\ \citenamefont {Vuleti\ifmmode~\acute{c}\else \'{c}\fi{}}}]{Hur2022}%
  \BibitemOpen
  \bibfield  {author} {\bibinfo {author} {\bibfnamefont {J.}~\bibnamefont {Hur}}, \bibinfo {author} {\bibfnamefont {D.~P.~L.}\ \bibnamefont {Aude~Craik}}, \bibinfo {author} {\bibfnamefont {I.}~\bibnamefont {Counts}}, \bibinfo {author} {\bibfnamefont {E.}~\bibnamefont {Knyazev}}, \bibinfo {author} {\bibfnamefont {L.}~\bibnamefont {Caldwell}}, \bibinfo {author} {\bibfnamefont {C.}~\bibnamefont {Leung}}, \bibinfo {author} {\bibfnamefont {S.}~\bibnamefont {Pandey}}, \bibinfo {author} {\bibfnamefont {J.~C.}\ \bibnamefont {Berengut}}, \bibinfo {author} {\bibfnamefont {A.}~\bibnamefont {Geddes}}, \bibinfo {author} {\bibfnamefont {W.}~\bibnamefont {Nazarewicz}}, \bibinfo {author} {\bibfnamefont {P.-G.}\ \bibnamefont {Reinhard}}, \bibinfo {author} {\bibfnamefont {A.}~\bibnamefont {Kawasaki}}, \bibinfo {author} {\bibfnamefont {H.}~\bibnamefont {Jeon}}, \bibinfo {author} {\bibfnamefont {W.}~\bibnamefont {Jhe}},\ and\ \bibinfo {author} {\bibfnamefont {V.}~\bibnamefont {Vuleti\ifmmode~\acute{c}\else \'{c}\fi{}}},\
  }\bibfield  {title} {\bibinfo {title} {{Evidence of two-source King plot nonlinearity in spectroscopic search for new boson}},\ }\href {https://doi.org/10.1103/PhysRevLett.128.163201} {\bibfield  {journal} {\bibinfo  {journal} {Phys. Rev. Lett.}\ }\textbf {\bibinfo {volume} {128}},\ \bibinfo {pages} {163201} (\bibinfo {year} {2022})}\BibitemShut {NoStop}%
\bibitem [{\citenamefont {Debierre}\ \emph {et~al.}(2022)\citenamefont {Debierre}, \citenamefont {Oreshkina}, \citenamefont {Valuev}, \citenamefont {Harman},\ and\ \citenamefont {Keitel}}]{Debierre2022}%
  \BibitemOpen
  \bibfield  {author} {\bibinfo {author} {\bibfnamefont {V.}~\bibnamefont {Debierre}}, \bibinfo {author} {\bibfnamefont {N.~S.}\ \bibnamefont {Oreshkina}}, \bibinfo {author} {\bibfnamefont {I.~A.}\ \bibnamefont {Valuev}}, \bibinfo {author} {\bibfnamefont {Z.}~\bibnamefont {Harman}},\ and\ \bibinfo {author} {\bibfnamefont {C.~H.}\ \bibnamefont {Keitel}},\ }\bibfield  {title} {\bibinfo {title} {Testing standard-model extensions with isotope shifts in few-electron ions},\ }\href {https://doi.org/10.1103/PhysRevA.106.062801} {\bibfield  {journal} {\bibinfo  {journal} {Phys. Rev. A}\ }\textbf {\bibinfo {volume} {106}},\ \bibinfo {pages} {062801} (\bibinfo {year} {2022})}\BibitemShut {NoStop}%
\bibitem [{\citenamefont {Witkowski}\ \emph {et~al.}(2019)\citenamefont {Witkowski}, \citenamefont {Kowzan}, \citenamefont {Munoz-Rodriguez}, \citenamefont {Ciury{\l}o}, \citenamefont {{\.Z}uchowski}, \citenamefont {Mas{\l}owski},\ and\ \citenamefont {Zawada}}]{Witkowski2019}%
  \BibitemOpen
  \bibfield  {author} {\bibinfo {author} {\bibfnamefont {M.}~\bibnamefont {Witkowski}}, \bibinfo {author} {\bibfnamefont {G.}~\bibnamefont {Kowzan}}, \bibinfo {author} {\bibfnamefont {R.}~\bibnamefont {Munoz-Rodriguez}}, \bibinfo {author} {\bibfnamefont {R.}~\bibnamefont {Ciury{\l}o}}, \bibinfo {author} {\bibfnamefont {P.~S.}\ \bibnamefont {{\.Z}uchowski}}, \bibinfo {author} {\bibfnamefont {P.}~\bibnamefont {Mas{\l}owski}},\ and\ \bibinfo {author} {\bibfnamefont {M.}~\bibnamefont {Zawada}},\ }\bibfield  {title} {\bibinfo {title} {{Absolute frequency and isotope shift measurements of mercury $^1S_0-^3P_1$ transition}},\ }\href@noop {} {\bibfield  {journal} {\bibinfo  {journal} {Optics Express}\ }\textbf {\bibinfo {volume} {27}},\ \bibinfo {pages} {11069} (\bibinfo {year} {2019})}\BibitemShut {NoStop}%
\bibitem [{\citenamefont {Safronova}\ \emph {et~al.}(2018)\citenamefont {Safronova}, \citenamefont {Budker}, \citenamefont {DeMille}, \citenamefont {Kimball}, \citenamefont {Derevianko},\ and\ \citenamefont {Clark}}]{Safronova2018}%
  \BibitemOpen
  \bibfield  {author} {\bibinfo {author} {\bibfnamefont {M.~S.}\ \bibnamefont {Safronova}}, \bibinfo {author} {\bibfnamefont {D.}~\bibnamefont {Budker}}, \bibinfo {author} {\bibfnamefont {D.}~\bibnamefont {DeMille}}, \bibinfo {author} {\bibfnamefont {D.~F.~J.}\ \bibnamefont {Kimball}}, \bibinfo {author} {\bibfnamefont {A.}~\bibnamefont {Derevianko}},\ and\ \bibinfo {author} {\bibfnamefont {C.~W.}\ \bibnamefont {Clark}},\ }\bibfield  {title} {\bibinfo {title} {Search for new physics with atoms and molecules},\ }\href {https://doi.org/10.1103/RevModPhys.90.025008} {\bibfield  {journal} {\bibinfo  {journal} {Rev. Mod. Phys.}\ }\textbf {\bibinfo {volume} {90}},\ \bibinfo {pages} {025008} (\bibinfo {year} {2018})}\BibitemShut {NoStop}%
\bibitem [{\citenamefont {McFerran}\ \emph {et~al.}(2012)\citenamefont {McFerran}, \citenamefont {Yi}, \citenamefont {Mejri}, \citenamefont {Di~Manno}, \citenamefont {Zhang}, \citenamefont {Gu\'ena}, \citenamefont {Le~Coq},\ and\ \citenamefont {Bize}}]{McFerran2012}%
  \BibitemOpen
  \bibfield  {author} {\bibinfo {author} {\bibfnamefont {J.~J.}\ \bibnamefont {McFerran}}, \bibinfo {author} {\bibfnamefont {L.}~\bibnamefont {Yi}}, \bibinfo {author} {\bibfnamefont {S.}~\bibnamefont {Mejri}}, \bibinfo {author} {\bibfnamefont {S.}~\bibnamefont {Di~Manno}}, \bibinfo {author} {\bibfnamefont {W.}~\bibnamefont {Zhang}}, \bibinfo {author} {\bibfnamefont {J.}~\bibnamefont {Gu\'ena}}, \bibinfo {author} {\bibfnamefont {Y.}~\bibnamefont {Le~Coq}},\ and\ \bibinfo {author} {\bibfnamefont {S.}~\bibnamefont {Bize}},\ }\bibfield  {title} {\bibinfo {title} {Neutral atom frequency reference in the deep ultraviolet with $\mathrm{\text{fractional uncertainty}}=5.7\ifmmode\times\else\texttimes\fi{}{10}^{\ensuremath{-}15}$},\ }\href {https://doi.org/10.1103/PhysRevLett.108.183004} {\bibfield  {journal} {\bibinfo  {journal} {Phys. Rev. Lett.}\ }\textbf {\bibinfo {volume} {108}},\ \bibinfo {pages} {183004} (\bibinfo {year} {2012})}\BibitemShut {NoStop}%
\bibitem [{\citenamefont {Graner}\ \emph {et~al.}(2016)\citenamefont {Graner}, \citenamefont {Chen}, \citenamefont {Lindahl}, \citenamefont {Heckel} \emph {et~al.}}]{Graner2016}%
  \BibitemOpen
  \bibfield  {author} {\bibinfo {author} {\bibfnamefont {B.}~\bibnamefont {Graner}}, \bibinfo {author} {\bibfnamefont {Y.}~\bibnamefont {Chen}}, \bibinfo {author} {\bibfnamefont {E.}~\bibnamefont {Lindahl}}, \bibinfo {author} {\bibfnamefont {B.}~\bibnamefont {Heckel}}, \emph {et~al.},\ }\bibfield  {title} {\bibinfo {title} {{Reduced limit on the permanent electric dipole moment of $^{199}$Hg}},\ }\href@noop {} {\bibfield  {journal} {\bibinfo  {journal} {Physical review letters}\ }\textbf {\bibinfo {volume} {116}},\ \bibinfo {pages} {161601} (\bibinfo {year} {2016})}\BibitemShut {NoStop}%
\bibitem [{\citenamefont {Angstmann}\ \emph {et~al.}(2004)\citenamefont {Angstmann}, \citenamefont {Dzuba},\ and\ \citenamefont {Flambaum}}]{Angstmann2004}%
  \BibitemOpen
  \bibfield  {author} {\bibinfo {author} {\bibfnamefont {E.~J.}\ \bibnamefont {Angstmann}}, \bibinfo {author} {\bibfnamefont {V.~A.}\ \bibnamefont {Dzuba}},\ and\ \bibinfo {author} {\bibfnamefont {V.~V.}\ \bibnamefont {Flambaum}},\ }\bibfield  {title} {\bibinfo {title} {Relativistic effects in two valence-electron atoms and ions and the search for variation of the fine-structure constant},\ }\href {https://doi.org/10.1103/PhysRevA.70.014102} {\bibfield  {journal} {\bibinfo  {journal} {Phys. Rev. A}\ }\textbf {\bibinfo {volume} {70}},\ \bibinfo {pages} {014102} (\bibinfo {year} {2004})}\BibitemShut {NoStop}%
\bibitem [{\citenamefont {Clivati}\ \emph {et~al.}(2020)\citenamefont {Clivati}, \citenamefont {Gravina}, \citenamefont {Castrillo}, \citenamefont {Costanzo}, \citenamefont {Levi},\ and\ \citenamefont {Gianfrani}}]{Clivati2020}%
  \BibitemOpen
  \bibfield  {author} {\bibinfo {author} {\bibfnamefont {C.}~\bibnamefont {Clivati}}, \bibinfo {author} {\bibfnamefont {S.}~\bibnamefont {Gravina}}, \bibinfo {author} {\bibfnamefont {A.}~\bibnamefont {Castrillo}}, \bibinfo {author} {\bibfnamefont {G.~A.}\ \bibnamefont {Costanzo}}, \bibinfo {author} {\bibfnamefont {F.}~\bibnamefont {Levi}},\ and\ \bibinfo {author} {\bibfnamefont {L.}~\bibnamefont {Gianfrani}},\ }\bibfield  {title} {\bibinfo {title} {{Tunable UV spectrometer for Doppler broadening thermometry of mercury}},\ }\href@noop {} {\bibfield  {journal} {\bibinfo  {journal} {Optics Letters}\ }\textbf {\bibinfo {volume} {45}},\ \bibinfo {pages} {3693} (\bibinfo {year} {2020})}\BibitemShut {NoStop}%
\bibitem [{\citenamefont {Gravina}\ \emph {et~al.}(2023)\citenamefont {Gravina}, \citenamefont {Clivati}, \citenamefont {Chishti}, \citenamefont {Castrillo}, \citenamefont {Fasci}, \citenamefont {Bertiglia}, \citenamefont {Lopardo}, \citenamefont {Sorgi}, \citenamefont {Coluccelli}, \citenamefont {Galzerano}, \citenamefont {Pastor}, \citenamefont {Levi},\ and\ \citenamefont {Gianfrani}}]{Gravina2023}%
  \BibitemOpen
  \bibfield  {author} {\bibinfo {author} {\bibfnamefont {S.}~\bibnamefont {Gravina}}, \bibinfo {author} {\bibfnamefont {C.}~\bibnamefont {Clivati}}, \bibinfo {author} {\bibfnamefont {N.~A.}\ \bibnamefont {Chishti}}, \bibinfo {author} {\bibfnamefont {A.}~\bibnamefont {Castrillo}}, \bibinfo {author} {\bibfnamefont {E.}~\bibnamefont {Fasci}}, \bibinfo {author} {\bibfnamefont {F.}~\bibnamefont {Bertiglia}}, \bibinfo {author} {\bibfnamefont {G.}~\bibnamefont {Lopardo}}, \bibinfo {author} {\bibfnamefont {A.}~\bibnamefont {Sorgi}}, \bibinfo {author} {\bibfnamefont {N.}~\bibnamefont {Coluccelli}}, \bibinfo {author} {\bibfnamefont {G.}~\bibnamefont {Galzerano}}, \bibinfo {author} {\bibfnamefont {P.~C.}\ \bibnamefont {Pastor}}, \bibinfo {author} {\bibfnamefont {F.}~\bibnamefont {Levi}},\ and\ \bibinfo {author} {\bibfnamefont {L.}~\bibnamefont {Gianfrani}},\ }\bibfield  {title} {\bibinfo {title} {Comb-assisted mercury spectroscopy in the deep-ultraviolet: towards the development of a new primary thermometer},\ }\href
  {https://doi.org/10.1088/1742-6596/2439/1/012015} {\bibfield  {journal} {\bibinfo  {journal} {Journal of Physics: Conference Series}\ }\textbf {\bibinfo {volume} {2439}},\ \bibinfo {pages} {012015} (\bibinfo {year} {2023})}\BibitemShut {NoStop}%
\bibitem [{\citenamefont {Machin}(2018)}]{Machin}%
  \BibitemOpen
  \bibfield  {author} {\bibinfo {author} {\bibfnamefont {G.}~\bibnamefont {Machin}},\ }\bibfield  {title} {\bibinfo {title} {The kelvin redefined},\ }\href {https://doi.org/10.1088/1361-6501/aa9ddb} {\bibfield  {journal} {\bibinfo  {journal} {Measurement Science and Technology}\ }\textbf {\bibinfo {volume} {29}},\ \bibinfo {pages} {022001} (\bibinfo {year} {2018})}\BibitemShut {NoStop}%
\bibitem [{\citenamefont {Gianfrani}(2016)}]{Gianfrani2016linking}%
  \BibitemOpen
  \bibfield  {author} {\bibinfo {author} {\bibfnamefont {L.}~\bibnamefont {Gianfrani}},\ }\bibfield  {title} {\bibinfo {title} {{Linking the thermodynamic temperature to an optical frequency: recent advances in Doppler broadening thermometry}},\ }\href@noop {} {\bibfield  {journal} {\bibinfo  {journal} {Philosophical Transactions of the Royal Society A: Mathematical, Physical and Engineering Sciences}\ }\textbf {\bibinfo {volume} {374}},\ \bibinfo {pages} {20150047} (\bibinfo {year} {2016})}\BibitemShut {NoStop}%
\bibitem [{\citenamefont {Gravina}\ \emph {et~al.}(2022)\citenamefont {Gravina}, \citenamefont {Clivati}, \citenamefont {Castrillo}, \citenamefont {Fasci}, \citenamefont {Chishti}, \citenamefont {Galzerano}, \citenamefont {Levi},\ and\ \citenamefont {Gianfrani}}]{Gravina2022}%
  \BibitemOpen
  \bibfield  {author} {\bibinfo {author} {\bibfnamefont {S.}~\bibnamefont {Gravina}}, \bibinfo {author} {\bibfnamefont {C.}~\bibnamefont {Clivati}}, \bibinfo {author} {\bibfnamefont {A.}~\bibnamefont {Castrillo}}, \bibinfo {author} {\bibfnamefont {E.}~\bibnamefont {Fasci}}, \bibinfo {author} {\bibfnamefont {N.}~\bibnamefont {Chishti}}, \bibinfo {author} {\bibfnamefont {G.}~\bibnamefont {Galzerano}}, \bibinfo {author} {\bibfnamefont {F.}~\bibnamefont {Levi}},\ and\ \bibinfo {author} {\bibfnamefont {L.}~\bibnamefont {Gianfrani}},\ }\bibfield  {title} {\bibinfo {title} {{Measurement of the mercury (6s6p) $^3$P$_1$-state lifetime in the frequency domain from integrated absorbance data}},\ }\href@noop {} {\bibfield  {journal} {\bibinfo  {journal} {Physical Review Research}\ }\textbf {\bibinfo {volume} {4}},\ \bibinfo {pages} {033240} (\bibinfo {year} {2022})}\BibitemShut {NoStop}%
\bibitem [{\citenamefont {Gambetta}\ \emph {et~al.}(2018)\citenamefont {Gambetta}, \citenamefont {Vicentini}, \citenamefont {Coluccelli}, \citenamefont {Wang}, \citenamefont {Fernandez}, \citenamefont {Maddaloni}, \citenamefont {De~Natale}, \citenamefont {Castrillo}, \citenamefont {Gianfrani}, \citenamefont {Laporta},\ and\ \citenamefont {Galzerano}}]{Gambetta}%
  \BibitemOpen
  \bibfield  {author} {\bibinfo {author} {\bibfnamefont {A.}~\bibnamefont {Gambetta}}, \bibinfo {author} {\bibfnamefont {E.}~\bibnamefont {Vicentini}}, \bibinfo {author} {\bibfnamefont {N.}~\bibnamefont {Coluccelli}}, \bibinfo {author} {\bibfnamefont {Y.}~\bibnamefont {Wang}}, \bibinfo {author} {\bibfnamefont {T.~T.}\ \bibnamefont {Fernandez}}, \bibinfo {author} {\bibfnamefont {P.}~\bibnamefont {Maddaloni}}, \bibinfo {author} {\bibfnamefont {P.}~\bibnamefont {De~Natale}}, \bibinfo {author} {\bibfnamefont {A.}~\bibnamefont {Castrillo}}, \bibinfo {author} {\bibfnamefont {L.}~\bibnamefont {Gianfrani}}, \bibinfo {author} {\bibfnamefont {P.}~\bibnamefont {Laporta}},\ and\ \bibinfo {author} {\bibfnamefont {G.}~\bibnamefont {Galzerano}},\ }\bibfield  {title} {\bibinfo {title} {{Versatile mid-infrared frequency-comb referenced sub-Doppler spectrometer}},\ }\href {https://doi.org/10.1063/1.5025135} {\bibfield  {journal} {\bibinfo  {journal} {APL Photonics}\ }\textbf {\bibinfo {volume} {3}},\ \bibinfo {pages} {046103}
  (\bibinfo {year} {2018})}\BibitemShut {NoStop}%
\bibitem [{\citenamefont {Westberg}\ \emph {et~al.}(2012)\citenamefont {Westberg}, \citenamefont {Wang},\ and\ \citenamefont {Axner}}]{Westberg2012}%
  \BibitemOpen
  \bibfield  {author} {\bibinfo {author} {\bibfnamefont {J.}~\bibnamefont {Westberg}}, \bibinfo {author} {\bibfnamefont {J.}~\bibnamefont {Wang}},\ and\ \bibinfo {author} {\bibfnamefont {O.}~\bibnamefont {Axner}},\ }\bibfield  {title} {\bibinfo {title} {{Fast and non-approximate methodology for calculation of wavelength-modulated Voigt lineshape functions suitable for real-time curve fitting}},\ }\href@noop {} {\bibfield  {journal} {\bibinfo  {journal} {Journal of Quantitative Spectroscopy and Radiative Transfer}\ }\textbf {\bibinfo {volume} {113}},\ \bibinfo {pages} {2049} (\bibinfo {year} {2012})}\BibitemShut {NoStop}%
\bibitem [{\citenamefont {Rohart}\ \emph {et~al.}(2014)\citenamefont {Rohart}, \citenamefont {Mejri}, \citenamefont {Sow}, \citenamefont {Tokunaga}, \citenamefont {Chardonnet}, \citenamefont {Darqui{\'e}}, \citenamefont {Dinesan}, \citenamefont {Fasci}, \citenamefont {Castrillo}, \citenamefont {Gianfrani} \emph {et~al.}}]{Rohart2014}%
  \BibitemOpen
  \bibfield  {author} {\bibinfo {author} {\bibfnamefont {F.}~\bibnamefont {Rohart}}, \bibinfo {author} {\bibfnamefont {S.}~\bibnamefont {Mejri}}, \bibinfo {author} {\bibfnamefont {P.~L.~T.}\ \bibnamefont {Sow}}, \bibinfo {author} {\bibfnamefont {S.~K.}\ \bibnamefont {Tokunaga}}, \bibinfo {author} {\bibfnamefont {C.}~\bibnamefont {Chardonnet}}, \bibinfo {author} {\bibfnamefont {B.}~\bibnamefont {Darqui{\'e}}}, \bibinfo {author} {\bibfnamefont {H.}~\bibnamefont {Dinesan}}, \bibinfo {author} {\bibfnamefont {E.}~\bibnamefont {Fasci}}, \bibinfo {author} {\bibfnamefont {A.}~\bibnamefont {Castrillo}}, \bibinfo {author} {\bibfnamefont {L.}~\bibnamefont {Gianfrani}}, \emph {et~al.},\ }\bibfield  {title} {\bibinfo {title} {{Absorption-line-shape recovery beyond the detection-bandwidth limit: application to the precision spectroscopic measurement of the Boltzmann constant}},\ }\href@noop {} {\bibfield  {journal} {\bibinfo  {journal} {Physical Review A}\ }\textbf {\bibinfo {volume} {90}},\ \bibinfo {pages} {042506}
  (\bibinfo {year} {2014})}\BibitemShut {NoStop}%
\bibitem [{\citenamefont {Rohart}(2017)}]{Rohart2017}%
  \BibitemOpen
  \bibfield  {author} {\bibinfo {author} {\bibfnamefont {F.}~\bibnamefont {Rohart}},\ }\bibfield  {title} {\bibinfo {title} {Overcoming the detection bandwidth limit in precision spectroscopy: The analytical apparatus function for a stepped frequency scan},\ }\href {https://doi.org/https://doi.org/10.1016/j.jqsrt.2016.09.019} {\bibfield  {journal} {\bibinfo  {journal} {Journal of Quantitative Spectroscopy and Radiative Transfer}\ }\textbf {\bibinfo {volume} {187}},\ \bibinfo {pages} {490} (\bibinfo {year} {2017})}\BibitemShut {NoStop}%
\bibitem [{\citenamefont {Stalnaker}\ \emph {et~al.}(2006)\citenamefont {Stalnaker}, \citenamefont {Budker}, \citenamefont {Freedman}, \citenamefont {Guzman}, \citenamefont {Rochester},\ and\ \citenamefont {Yashchuk}}]{Budker2006}%
  \BibitemOpen
  \bibfield  {author} {\bibinfo {author} {\bibfnamefont {J.~E.}\ \bibnamefont {Stalnaker}}, \bibinfo {author} {\bibfnamefont {D.}~\bibnamefont {Budker}}, \bibinfo {author} {\bibfnamefont {S.~J.}\ \bibnamefont {Freedman}}, \bibinfo {author} {\bibfnamefont {J.~S.}\ \bibnamefont {Guzman}}, \bibinfo {author} {\bibfnamefont {S.~M.}\ \bibnamefont {Rochester}},\ and\ \bibinfo {author} {\bibfnamefont {V.~V.}\ \bibnamefont {Yashchuk}},\ }\bibfield  {title} {\bibinfo {title} {{Dynamic Stark effect and forbidden-transition spectral line shapes}},\ }\href {https://doi.org/10.1103/PhysRevA.73.043416} {\bibfield  {journal} {\bibinfo  {journal} {Phys. Rev. A}\ }\textbf {\bibinfo {volume} {73}},\ \bibinfo {pages} {043416} (\bibinfo {year} {2006})}\BibitemShut {NoStop}%
\bibitem [{\citenamefont {Lewis}(1980)}]{Lewis1980}%
  \BibitemOpen
  \bibfield  {author} {\bibinfo {author} {\bibfnamefont {E.}~\bibnamefont {Lewis}},\ }\bibfield  {title} {\bibinfo {title} {Collisional relaxation of atomic excited states, line broadening and interatomic interactions},\ }\href {https://doi.org/https://doi.org/10.1016/0370-1573(80)90056-3} {\bibfield  {journal} {\bibinfo  {journal} {Physics Reports}\ }\textbf {\bibinfo {volume} {58}},\ \bibinfo {pages} {1} (\bibinfo {year} {1980})}\BibitemShut {NoStop}%
\bibitem [{\citenamefont {Petersen}\ \emph {et~al.}(2008)\citenamefont {Petersen}, \citenamefont {Chicireanu}, \citenamefont {Dawkins}, \citenamefont {Magalh\~aes}, \citenamefont {Mandache}, \citenamefont {Le~Coq}, \citenamefont {Clairon},\ and\ \citenamefont {Bize}}]{Petersen}%
  \BibitemOpen
  \bibfield  {author} {\bibinfo {author} {\bibfnamefont {M.}~\bibnamefont {Petersen}}, \bibinfo {author} {\bibfnamefont {R.}~\bibnamefont {Chicireanu}}, \bibinfo {author} {\bibfnamefont {S.~T.}\ \bibnamefont {Dawkins}}, \bibinfo {author} {\bibfnamefont {D.~V.}\ \bibnamefont {Magalh\~aes}}, \bibinfo {author} {\bibfnamefont {C.}~\bibnamefont {Mandache}}, \bibinfo {author} {\bibfnamefont {Y.}~\bibnamefont {Le~Coq}}, \bibinfo {author} {\bibfnamefont {A.}~\bibnamefont {Clairon}},\ and\ \bibinfo {author} {\bibfnamefont {S.}~\bibnamefont {Bize}},\ }\bibfield  {title} {\bibinfo {title} {{Doppler-free spectroscopy of the $^{1}S_{0}\mathrm{\text{\ensuremath{-}}}^{3}P_{0}$ optical clock transition in laser-cooled fermionic isotopes of neutral mercury}},\ }\href {https://doi.org/10.1103/PhysRevLett.101.183004} {\bibfield  {journal} {\bibinfo  {journal} {Phys. Rev. Lett.}\ }\textbf {\bibinfo {volume} {101}},\ \bibinfo {pages} {183004} (\bibinfo {year} {2008})}\BibitemShut {NoStop}%
\bibitem [{\citenamefont {Schelfhout}\ and\ \citenamefont {McFerran}(2021)}]{McFerran}%
  \BibitemOpen
  \bibfield  {author} {\bibinfo {author} {\bibfnamefont {J.~S.}\ \bibnamefont {Schelfhout}}\ and\ \bibinfo {author} {\bibfnamefont {J.~J.}\ \bibnamefont {McFerran}},\ }\bibfield  {title} {\bibinfo {title} {{Isotope shifts for $^{1}S_{0}\ensuremath{-}^{3}P_{0,1}^{o}$ Yb lines from multiconfiguration Dirac-Hartree-Fock calculations}},\ }\href {https://doi.org/10.1103/PhysRevA.104.022806} {\bibfield  {journal} {\bibinfo  {journal} {Phys. Rev. A}\ }\textbf {\bibinfo {volume} {104}},\ \bibinfo {pages} {022806} (\bibinfo {year} {2021})}\BibitemShut {NoStop}%
\bibitem [{\citenamefont {des Poids~et Mesures~(BIPM)}(2013)}]{BIPM}%
  \BibitemOpen
  \bibfield  {author} {\bibinfo {author} {\bibnamefont {Bureau International des Poids~et Mesures~(BIPM)}},\ }\bibfield  {title} {\bibinfo {title} {{Estimation of a consensus KCRV and associated degrees of equivalence}},\ }\bibfield  {journal} {\bibinfo  {journal} {CCQM Guidance note}\ }\href {https://doi.org/https://www.bipm.org/documents/20126/28430045/working-document-ID-5794/49d366bc-295f-18ca-c4d3-d68aa54077b5}{https://www.bipm.org/documents/20126/28430045/working-document-ID-5794/49d366bc-295f-18ca-c4d3-d68aa54077b5} (\bibinfo {year} {2013})\BibitemShut {NoStop}%
\end{thebibliography}
\providecommand{\noopsort}[1]{}\providecommand{\singleletter}[1]{#1}%

\end{document}


\preprint{APS/123-QED}

\title{Supplemental material for  ``Comb-referenced Doppler-free spectrometry of the $^{200}$Hg and $^{202}$Hg intercombination line at 254 nm"}
\author{Stefania Gravina}
 \affiliation{Dipartimento di Matematica e Fisica, Università degli Studi della Campania "Luigi Vanvitelli", 81100, Caserta, Italia}
\author{Naveed A. Chishti}%
\affiliation{%
Dipartimento di Matematica e Fisica, Università degli Studi della Campania "Luigi Vanvitelli", 81100, Caserta, Italia}%
\author{Simona Di Bernardo}
\affiliation{%
Dipartimento di Matematica e Fisica, Università degli Studi della Campania "Luigi Vanvitelli", 81100, Caserta, Italia}%
\author{Eugenio Fasci}
\affiliation{Dipartimento di Matematica e Fisica, Università degli Studi della Campania "Luigi Vanvitelli", 81100, Caserta, Italia}%
\author{Antonio Castrillo}
\affiliation{Dipartimento di Matematica e Fisica, Università degli Studi della Campania "Luigi Vanvitelli", 81100, Caserta, Italia}%
\author{Athanasios Laliotis}
\affiliation{Laboratoire de Physique des Lasers, Université Sorbonne Paris Nord, 93430, Villetaneuse, France}
\author{Livio Gianfrani}
\affiliation{%
Dipartimento di Matematica e Fisica, Università degli Studi della Campania "Luigi Vanvitelli", 81100, Caserta, Italia}%

\date{\today}

\begin{abstract}

\end{abstract}

\maketitle


\section{The comb-locking approach}
The external-cavity diode laser (ECDL) at 1014.8 nm is tighly locked to an optical frequency comb synthesizer (MENLO Systems, FC1500-250-WG) based on a mode-locked, femtosecond, erbium-doped fiber laser at 1560 nm. The comb repetition rate ($f_{rep}$) and the carrier-envelope offset frequency ($f_{ceo}$) are meticulously stabilized against a GPS-disciplined Rb clock. This configuration provides a self-referenced supercontinuum that covers one octave at wavelengths from 1050 nm to 2100 nm.
Since the ECDL wavelength falls outside this range, comb locking required the implementation of a novel scheme based on nonlinear mixing.
Very briefly, frequency doubling of
the longer wavelength portion of the comb (above 2 $\mu$m) in a 3-mm-long MgO-doped periodically poled lithium niobate
crystal was performed to produce a mini-comb centered at 1020 nm with a width of about 20 nm and a peak power of 1 $\mu$W.
The beat note between the ECDL and the nearest tooth of the mini-comb was observed by using a fast photodiode and compared with a local oscillator (LO), which in turn is phase locked to the Rb clock. The error signal is then preamplified, integrated, and fed back to the diode laser current so as to achieve a tight lock of the ECDL frequency.

\section{The lineshape model}
In the limit of a negligible intensity modulation associated to the wavelength modulation at $f_m$, the in-phase component of the n$^{th}$ harmonic output of the lock-in amplifier is proportional to the even component of the n$^{th}$ Fourier coefficient of the detector signal.  
Therefore, for a modulated Voigt absorption lineshape in the Doppler-limited linear regime, the in-phase lock-in signal, in coincidence with the n$^{th}$ harmonic detection, is given by the following expression:
\begin{equation}
S_{V,n}^{in} =-\beta I_{pr} \times \chi_{V,n}^{even}(\bar{\nu}_d, \bar{\nu}_a),
\label{lockin}
\end{equation}
where $I_{pr}$ is the intensity of the probe laser beam and $\beta$ is a proportionality factor that includes the gain of the lock-in amplifier, the detector sensitivity, and the frequency-integrated line absorbance. It is worth noting that the even component of the n$^{th}$ Fourier coefficient of the wavelength-modulated Voigt convolution, $\chi_{V,n}^{even}$, depends on $\bar{\nu}_d$ and $\bar{\nu}_a$ that are the width-normalized detuning frequency and modulation depth, given by $(\nu-\nu_0)/\delta\nu_L$ and
$\nu_a/\delta\nu_L$, respectively; here $\nu$ represents the laser frequency, $\nu_{0}$ is the line center frequency, $\nu_{a}$ is the modulation depth, and $\delta\nu_L$ is the half-width at half-maximum (HWHM) of the Lorentzian profile. 

The extension of this model to wavelength-modulated Lamb-dip profiles is straightforward. Under the 1$^{st}$ harmonic detection regime ($n$=1), the in-phase lock-in signal is given by:
\begin{equation}
S_{V,1}^{in}(\nu-\nu_0)=P_0+P_1 G_D(\nu-\nu_0)+P_2 \chi_{V,1}^{even}(\bar{\nu}_d, \bar{\nu}_a),
\end{equation}
where $P_0$ is a baseline parameter, while $P_1$ and $P_2$ are two amplitude parameters, the first one giving the magnitude of the residual Doppler-broadened background, which is given by the first derivative ($G_D$) of a Gaussian function. As for $P_2$, it plays the same role of the product $\beta$$I_{pr}$ of Eq. \eqref{lockin}. 

According to the Westberg-Wang-Axner methodology, the even component of the n$^{th}$ Fourier coefficient
of the wavelength-modulated Voigt convolution can be written as follows:
\begin{eqnarray}
&&\chi_{V,n}^{even}(\bar{\nu}_d, \bar{\nu}_a) = \dfrac{y}{\pi^{3/2}\delta\nu_L}\times\nonumber\\
&&\int^{+\infty}_{-\infty}\dfrac{2}{\tau_D}
\int^{\tau_D}_0\dfrac{\cos(2\pi n f_m t)}{1+[(\bar{\nu}_d+s')+\bar{\nu}_a \cos(2\pi n f_m t)]^2}\textnormal{dt}\nonumber\\
&& e^{-y^2s'^2}\textnormal{ds'},
\label{even}
\end{eqnarray}
where $y$ is the ratio of the Lorentzian and Gaussian widths multiplied by $\sqrt{\textnormal{ln2}}$ and $\tau_D$ is the lock-in integration time, which is supposed to be much larger than the inverse of the modulation frequency (as for our experimental conditions). Since the inner integral (over time) can be identified with the even component of the n$^{th}$ Fourier coefficient of the wavelength-modulated Lorentzian absorption lineshape function, which has an analytical expression, Eq. \eqref{even} can be simplified to a single integral, given by:
\begin{eqnarray}
&&\chi_{V,n}^{even}(\bar{\nu}_d, \bar{\nu}_a) =\dfrac{y}{\pi^{1/2}}\times\nonumber\\
&&\int^{+\infty}_{-\infty}\chi_{L,n}^{even}(\bar{\nu}_d+s', \bar{\nu}_a)  e^{-y^2s'^2}\textnormal{ds'}.
\end{eqnarray}
This latter takes the form of a convolution that can be effectively and accurately computed within the MATLAB environment.

The experimental lineshapes are compared to the theoretical profile by means of a nonlinear least-squares fitting procedure.
In the spectral analysis,  $P_0$, $P_1$, and $P_2$ are considered as free parameters, along with the line center frequency, the Gaussian and Lorentzian components of the Lamb-dip width, and the modulation depth. Instead, the width of the Doppler-broadened background is set to the value that corresponds to the gas temperature.